\documentstyle[sprocl]{article}
\def\Journal#1#2#3#4{{#1} {\bf #2}, #3 (#4)}

\def\NPB{{\em Nucl. Phys.} B}
\def\PLB{{\em Phys. Lett.}  B}
\def\PRL{\em Phys. Rev. Lett.}
\def\PRD{{\em Phys. Rev.} D}

\def\be{\begin{equation}}
\def\ee{\end{equation}}
\def\beq{\be}
\def\eeq{\ee}
\def\bea{\begin{eqnarray}}
\def\eea{\end{eqnarray}}

\begin{document}
\title{ Supersymmetric Partners
of Oblique Corrections~\footnote{Talk  presented by L. Randall at the SUSY97 Conference, May 27-31, Philadelphia PA, USA.}}
\author{ L. Randall,  E. Katz  and S. Su }
\address{ Center for Theoretical Physics, \\  
Laboratory for Nuclear Science and Department 
of Physics \\ Massachusetts Institute of Technology \\
 Cambridge, MA ~~ 02139, USA}
\maketitle\abstracts{ We discuss a potential new probe
 of supersymmetric physics.  In particular,
we discuss the possibility of  measuring hard supersymmetry
violation which occurs at one loop through ``super-oblique" corrections
to the gauge and gaugino propagators. In models
with heavy scalar partners, or with many gauge-charged particles
which participate in supersymmetry breaking, these effects
can be substantial due to logarithmic and multiplicity factor
enhancements.}

 \section{Introduction}
Supersymmetry has yet to be discovered.
 If/When it is, the information about the underlying theory
will be limited but powerful. 
 For example, the spectrum of superpartners might
tell us about the mechanism of communication of supersymmetry
breaking or consistency with a specific high-energy
model (GUT theory for example).
 However, these measurements are (probably) a long way off.
 In the meantime, it is useful to be prepared
to interpret what will be measured. This might
be through new insights into the fundamental underlying theory,
new models of the effective low-energy physics, or by
new ideas for interesting quantities to measure, and how
to interpret them.

In this talk, I will discuss a new category of precision measurements
which one might use to probe supersymmetric theories. As we
will discuss, these measurements have much in common with the
well-studied ``oblique'' corrections which characterized
corrections to the standard model. In this case, the measurements
will be more difficult, but should they prove possible,
could potentially yield invaluable insights into the underlying
physics. The work we discuss today will be presented
in more detail in a future publication \cite{us2}.

The key to understanding the quantities we wish to consider
is to distinguish hard and soft supersymmetry breaking parameters.
 The focus of most research on supersymmetric phenomenology
focusses on the soft supersymmetry breaking parameters,
in particular the superpartner masses. Hard supersymmetry
breaking, on the other hand, is ``forbidden''.
However, this is not true, even in the softly broken
supersymmetric theory! 
 The only constraint is that the  coefficients of hard supersymmetry
breaking operators are {\it finite}.
 There are no available counterterms in the supersymmetric Lagrangian.
 Nonetheless, finite corrections are permissible and
present.

 Our focus here is on new
parameters which reflect hard supersymmetry breaking
in the form of oblique corrections in the gauge sector,
akin to precision electroweak parameters.
Recall the merits of precision electroweak paramters. They
are 
 new parameters which reflect the spontaneous breaking of
the elcectroweak gauge symmetry.
They served as tests of the standard model,
and gave insights into   non-standard model physics.
Similarly,  new SUSY-oblique corrections will test
the MSSM, and serve as 
  probes of new physics.

\section{Oblique and Super-Oblique}

It is therefore useful to review standard oblique
corrections.
 Precision electroweak corrections were important
in that they 
 tested consistency of the standard model and
 gave insight into the high-energy world.
The oblique corrections
are the corrections to the gauge boson propagator.
The parameters which characterize
the most important effects can be obtained by retaining
the leading pieces in a derivative expansion,
accounting for gauge invariance. The six ``parameters''
are
 $\Pi'_{\gamma \gamma}$, $\Pi'_{\gamma Z}$, $\Pi_{WW}$,
$\Pi'_{WW}$, $\Pi_{ZZ}$, $\Pi'_{ZZ}$.
We chose to work in the basis of tree-level mass eigenstates
and  retained only the nondecoupling pieces
of the propagator, those of dimension four or less.
These are not the six parameters which were actually used
when comparing to measurements; it is more useful first
to absorb three of the parameters into
 $g$, $g'$, $v$.
 We are left with three new finite parameters which are forbidden
by gauge invariance but permitted once it is spontaneously broken.
With my collaborator Mitch Golden, we identified these three
parameters as \cite{gr}
 $\delta Z_{ZZ}$, $\delta Z_{\gamma Z}$, $\delta m_{ZZ}$.
In the more widely utilized Peskin-Takeuchi
naming convention, they were absorbed into parameters
\cite{pt1,pt2}  $S$, $T$, $U$, while a third popular convention
from a paper of Altarelli and Barbieri \cite{ab}
absorbed them in the three parameters
 $\epsilon_1$,
$\epsilon_2$, $\epsilon_3$.
The importance of these parameters is
that even heavy states do not decouple.
For example,  one can compute the contributions of a heavy
doublet to the $S$ parameter, $S=-4 \pi {d \over d q^2} \Pi_{3S}(q^2)|_{q^2=0}$
\beq
\Delta S={1 \over 6 \pi} \left[1-Y log ({m_N^2 \over m_E^2}) \right] .
\eeq

Of course, the many new particles present in a supersymmetric
theory can contribute to the standard oblique corrections.
In general, for reasonable values of the supersymmetric
partner masses, these effects are quite small. However,
these are not what we wish to discuss here.
We instead focus on  a  new class of
oblique corrections which one can define in supersymmetric theories.
 Supersymmetry guarantees equality of the gauge
and gaugino Yukawa couplings.
 When supersymmetry is broken, these
are no longer guaranteed to be equal.
It is easy to calculate the one-loop effect of a single
heavy doublet superfield to the difference of SU(2) gauge
and SU(2) gaugino Yukawa coupling by calculating the
``super-oblique'' correction to the gauge propagator and gaugino
propagator. One thereby derives
 \[
{g-\tilde{g} \over g}={\alpha \over 24 \pi}\left( Log\left({ m_{\tilde{W}} \over m_{\tilde{q}}} \right)^2-{11 \over 12}\right)
\]
Here we have computed the contribution of a heavy doublet
squark to the difference in SU(2) gauge vs. gaugino coupling.
For physical quarks, they are generally lighter than the
gaugino, so we have cutoff the logarithm at the gaugino mass.
We have also evaluated the gauge coupling at the same momentum
scale as the gaugino mass.

 Although this is one-loop (and there is an  additional 1/6 suppression),
there are factors which can make it large. First of all,
we see that there is a logarithm; 
 if there is a large mass splitting, there will be  a corresonding
 enhancement.
Second, there is a multiplicity factor;
 ``every" particle with gauge charge contributes.
This last fact was also what contributed
to the importance of standard electroweak parameters.
However, the logarithm is different
here because decoupling works  differently than for standard
electroweak corrections. Heavy scalar superpartners contribute
{\it more} rather than less.

If we call the soft scalar mass $m_0$ and the coefficient of
the trilinear scalar coupling $A$, for the standard oblique
corrections (eg the $S$ parameter), it is easy to see that
\[
\Delta S \propto {m_{LR}^2 \over m^2}\sim {(\lambda_q A m_q )^2\over m_0^2} 
\]
which decouples for large $m_0$.

In the case of the SUSY-oblique corrections which we discuss,
one finds
\[
\delta g(k) \propto Log{m_{\tilde{q}}^2 \over  m_q^2}
\]
(where we now assume the quark mass is bigger than the gaugino's)
which clearly does not decouple, and is in fact
logarithmically enhanced.  This logarithm can however be readily understood \cite{nojiri}.
If we assume a chiral multiplet in which the scalar partner
is heavy, there is an energy regime between the fermion (or gaugino)
mass and the heavy scalar mass in which the gauge 
and gaugino coupling run differently.
 In fact one can also determine that when the scalar is the heavy
partner the gauge coupling at low energy will be reduced relative
to the gaugino couplings due to the sign of the fermion contribution
to the beta-function for the gauge coupling.

Decoupling does occur  however for SUSY-oblique
corrections when $m_q\sim m_{\tilde{q}}$
\[
\delta g \propto {(m_{\tilde{q}}^2-m_q^2) \over m_q^2}
\]
Notice here that  we have assumed   the squark and quark to be
  nearly degenerate and heavier than the gaugino.
Therefore what is relevant to SUSY-oblique corrections
are standard model gauge charge multiplets with large
supersymmetry breaking splitting in the masses.
We conclude that heavy
 scalar partners can lead to interesting deviations
from supersymmetric prediction $\delta g=0$.

\section{Large Effects and FCNC}

It is reasonable to ask whether there is any reason to believe
that there might be large mass splittings, or nonstandard model
gauged states. The answer might be yes, given the current
ideas for resolving the flavor-changing-neutral-current (FCNC)
problems in supersymmetric theories.
 The apparent degeneracy of squarks or alignment
with quarks is still not understood.
Potential explanations in the literature include gauge-mediated models
and symmetries. It has also been pointed out \cite{ckn,dp,dg}
   that  the problem is ameliorated
in models in which the first two generations of squarks are heavy.
One example of a class of theories for which this is the case
is the ``More Minimal Supersymmetric Standard Model'',
in which one attempts to allow the maximal masses consistent
with naturalness bounds.  This permits the first two generations
of squarks to be heavier which goes a considerable way towards
solving flavor changing problems.
One can then imagine
 $\tilde{m}_{1,2}\sim 20 {\rm TeV}$, $\tilde{m}_3 \sim 100{\rm GeV}-1 {\rm TeV}$.

A previous paper had also incorporated the idea that the first
two generations are heavy. Specifically, they incorporated
an anomalous U(1) symmetry as a mediator of supersymmetry
breaking, and imagined that the first two generations were
charged whereas the third generation was not.
 Squarks which are charged under $U(1)$ group
can have mass of order $4 \pi$ times as big as
usual supergravity induced mass because they get their mass
from the larger $D$-term.

Another example in which there might be large effects
are some models of gauge-mediation.  Gauge mediation models
can readily produce heavier squarks than gauginos
since a gaugino mass requires $R$-symmetry breaking.
An example of this was a class of models called ``Mediator Models'' \cite{mi}.
The idea behind these models is that
there is a  weakly gauged global symmetry $G_m$ in the dynamical
supersymmetry breaking sector and a SUSY breaking gaugino mass $M_m$.
In addition, there are ``mediator'' fields which communicate with
both sectors; specifically, one can have a vector-representation of
both groups $T(m,5)$, $\bar{T}(\bar{m},\bar{5})$ with a mass
 $M_T T \bar{T}$ .
 These models have the advantage that there are
no singlets, no complicated superpotential,
and the correct vacuum is global minimum.
 However, gaugino masses arise at three loops
whereas squark masses arise at two.
The generic prediction then 
is that   $m_{\tilde{q}}\sim 10 m_{1/2}$. This imposes naturalness
problems but can yield interesting experimental signatures so
it is testable.

A final example in which effects might be large is if there
is a messenger sector without great mass splitting but
with a large representation. That is, models
in which there are many objects carrying the gauge charge
can also yield big effects.

\section{Parameters and Numbers}

One can now ask for the  kinds of models described above, what
is the magnitude of the effects on gauge vs. gaugino coupling. 
  There are three measureable quantities associated
with $g_3$, $g_2$, $g_1$. These fit in well with a modified
Peskin-Takeuchi naming convention
 $v=(g_3-\tilde{g_3})/g_3$,
 $w=(g_2-\tilde{g_2})/g_2$,
 $y=(g_1-\tilde{g_1})/g_1$.
Of course, one can use alternative names
such as $\epsilon_4$,$\epsilon_5$, $\epsilon_6$.

We now calculate the numbers for each of these quantities
for the models described above in which there is a large mass splitting.
In reality, when the splittings are described as above in the
gauge mediated case, one needs to be careful to run the parameters.
We make the simple assumptions below to get an idea of the magnitude
of the effects.

We define a mediator model spectrum by
 $m_{\tilde{l}}=10 m_{\tilde{w}}$,
$m_{\tilde{q}}=20 m_{\tilde{w}}$. For this spectrum, we find 
$v=-10.5 \%$, $w=-3.4 \%$, $y=-1.7 \%$.

If we assume that sleptons are light and it is only the squarks
which are heavy (probably necessary for these quantities to
be measurable), one would calculate
 $w=-2.7 \%$, $y=-1.6 \%$.

We define a More Minimal Spectrum by
$b_R$, $u$, $d$, $s$, $c$, 
$e$, $\mu$, $H$ having mass 20 TeV, 
whereas $b_L$, $t$, $\tau$: 1 TeV, $m_{\tilde{w }}=100 {\rm GeV}$, $m_{\tilde{g}}=200{\rm GeV}$.
In this case, one obtains
$v=-16 \%$, $w=-5 \%$, $y=2.6 \%$.
For a modified more minimal model in which the sleptons have
mass 1 TeV instead, one obtains
$w=-4.8 \%$, $y=-2.1\%$.

\section{How to Measure?}

It is clear that there can be fairly sizable effects. It is therefore
extremely interesting to ask the question whether these parameters are
conceivably measurable. These numbers serve as benchmarks
for interesting measurements, although a measurement at any
level of accuracy will constrain our confidence in supersymmetry
or equivalently constrain extensions of the minimal scenario.

Several groups have begun to investigate this question.
Initially this was pursued as a test of supersymmetry at
tree level \cite{feng}
but it was later pursued also in order to extract squark masses
if they are heavy \cite{nojiri}.

The most precise measurement which has been suggested
is of the U(1) coupling in a paper by Nojiri, Fujii, and Tsukamoto
\cite{nojiri} where they studied
 $\sigma(e^+ e^- \to \tilde{e}_R^+ \tilde{e}_R^-)$.
This process
proceeds through $s$-channel gauge exchange and $t$-channel neutralino
exchange. One can measure
$\tilde{g}_1$ and  $M_1$ by measuring
the differential cross section
  ${d\sigma(e^+ e^- \to \tilde{e}_R^+ \tilde{e}_R^-) \over d \cos\theta}$
with accuracy at 1 \% level.
It has also been suggested by Cheng, Feng, and Polonsky to look
at the $e^- e^-$ scattering cross section when running the collider
in the appropriate mode, which can lead to even greater precision
\cite{feng,feng2}.  It should be borne in mind that
although the U(1) coupling can be most accurately measured,
the magnitude of $y$ is the smallest of the three precision
measurements. It is therefore worthwhile to investigate
the possibility of measuring $w$ and $v$ as well.

Two possible ways to measure $w$ have been suggested \cite{feng}.
If both 
  charginos can
be produced at the NLC, the gaugino coupling
can be tested through the chargino mass dependence
on the $H \tilde{H} \tilde{W}$ vertex. This only
permitted about a 
 15-30 \%  measurement.
 In the ``gaugino" region, where charginos are
very nearly pure gauginos, one can test the $e \tilde{\nu} \tilde{W}$
vertex directly and
  $w$ can be measured to about 15 \%. However, with
 independent measurement of $m_{\tilde{\nu}}$,
$w$ can be measured to better than  5 \%.

Finally we consider the measurement of $v$, which
requires measuring the SU(3) gaugino Yukawa coupling.
It is  very likely that  the best test will be at a hadron collider.
 At an $e^+ e^-$ collider, it might be possible
to test the MMSSM if  $\tilde{t}$, $\tilde{g}$ are sufficiently
light \footnote{We thank Michael Peskin for suggesting this
possibility} or if $\tilde{b}$, $\tilde{g}$ are sufficiently
light \footnote{We thank Jonathan Feng for pointing out this
alternative possibility.}  One can then measure the
 cross section for $t \tilde{t} \tilde{g}$ production or
$b \tilde{b} \tilde{g}$ production.  
For $b$ quark and squark production one predicts
a few hundred events according to the gluino and squark
mass (below  a few hundred GeV) and roughly half
this number  for tops. Details will be presented in  a future
paper \cite{us2}.
 Especially in the  first case, but possibly in the  second,
it is also possible that the masses are such that
the intermediate squark state is on shell, in which
case one is in fact measuring the branching fraction
into quark and gluino, which can also determine the
gaugino  coupling.  The number of events we presented
here is the production cross section but it  is not corrected
for cuts necessary to remove background events. Nonetheless
for sufficiently light gaugino and third generation squarks,
it seems likely that a measurement of the gaugino coupling
at at least the 15~ \% level should be possible. 

\section{More Non-Standard Physics}

Before concluding, we wish to stress the role of these
measurements as not only  a test of  the parameters of
the minimal extension of the standard model 
but also as an exciting probe of new physics.  We have
already stressed the fact that heavy squarks will lead
to larger effects.  This is especially important
as it is precisely when the squarks become kinematically
inaccessible that their virtual effects become largest.
If one does not find the squarks, or if they are at the
edge of the kinematically accessible regime, one would
like to be able to confirm their existence and attempt
some estimate of their mass through their virtual effects.

However, it might be that the squarks are relatively light,
and one still measures large deviations from the supersymmetric
relations. Provided they are not too large, one might
ask how this could be consistent with supersymmetry.
In fact, this is probably the most exciting possibility.
It would indicate the existence of
nonstandard fields which participate in supersymmetry
breaking and therefore have a large mass splitting relative
to the mass of the fermions or scalars and which carry
standard model gauge charge. An obvious candidate
in terms of existing models is a messenger sector,
although possibilities extend beyond this.
As an example, suppose the messenger sector consisted
of a  large vectorlike representatin under SU(5). One should
note that for the ``standard'' messenger scenario,
with a singlet coupled to messengers through a Yukawa
term,  the
masses of the scalars are $\lambda^2 S^2\pm \lambda F_S$ whereas the fermion
mass is $\lambda S$. It is readily checked in this case that
the logarithm takes a sign opposite to that before.  We note that
we need a nontrivial mass splitting in this nonstandard sector
to get a sizable contribution. This would indicate a strong
participation of this nonstandard model sector in supersymmetry
breaking and would give great insight into the high-energy world.

\section{Conclusions}
  
To conclude, it is obvious that we would like to gain
 as many handles into the
underlying supersymmetric theory as can be accessible.
 SUSY-Oblique corrections provide a different
perspective into the physics of supersymmetry breaking
and
 into the consistency of the standard supersymmetric sector.
In nonstandard models, especially those addressing
the supersymmetric flavor problem, it is likely
that the parameters we have described can be large.
Although it is almost certain that they will be a challenge
to measure at a very interesting level.
Clearly, the more precisely they are measured, the more
constrained will be the underlying physics and our
confidence in the supersymmetric model which is revealed through
mass measurements. It is definitely worthy of
further investigation.

\section*{Acknowledgments} We thank
Jonathan Feng, Ian Hinchliffe, and Michael
Peskin for conversations about measuring $v$.
 This research is supported in part by DOE under
cooperative agreement \#DE-FC02-94ER40818, NSF Young
Investigator Award, Alfred P. Sloan Foundation Fellowship,
DOE Outstanding Junior Investigator Award.

While presenting this work, we learned of related research 
by Cheng, Feng, Polonsky, and Pierce, Nojiri, and Yamada \cite{feng2,pierce}.

\end{document}